\documentclass[prl,showpacs,twocolumn]{revtex4}
\usepackage{graphicx}
\begin{document}
\title{Dissociation of ultracold molecules with Feshbach resonances}
\author{Stephan D{\"u}rr, Thomas Volz, and Gerhard Rempe}
\affiliation{Max-Planck-Institut f{\"u}r Quantenoptik, Hans-Kopfermann-Stra{\ss}e 1, 85748 Garching, Germany}
\date{\today}
\hyphenation{Fesh-bach}
\begin{abstract}
Ultracold molecules are associated from an atomic Bose-Einstein condensate by ramping a magnetic field across a Feshbach resonance. The reverse ramp dissociates the molecules. The kinetic energy released in the dissociation process is used to measure the widths of 4 Feshbach resonances in $^{87}$Rb. This method to determine the width works remarkably well for narrow resonances even in the presence of significant magnetic-field noise. In addition, a quasi-mono-energetic atomic wave is created by jumping the magnetic field across the Feshbach resonance.
\end{abstract}
%
\maketitle

The field of ultracold molecules has seen impressive progress over the course of the last year. A recent landmark achievement was the creation of a Bose-Einstein condensate (BEC) of molecules \cite{greiner:03,jochim:03,zwierlein:03,bourdel:cond-mat/0403091}. Presently, several experiments are exploring the crossover regime between BEC and BCS (Bardeen-Cooper-Schrieffer) superfluidity \cite{regal:04,bartenstein:04,zwierlein:04,bourdel:cond-mat/0403091,kinast:04}. While these experiments associate bosonic dimers from fermionic atoms (see also Ref.~\cite{strecker:03}), there is also considerable interest in dimers associated from bosonic atoms \cite{donley:02,herbig:03,duerr:04,xu:03}. Ultracold molecules are created by first cooling atoms to ultracold temperatures and then associating them to molecules using either photoassociation or a Feshbach resonance. In the latter case, the molecules are typically created by slowly ramping the magnetic field in the appropriate direction across the Feshbach resonance.

In this paper, we investigate the dissociation of molecules by ramping the magnetic field back through the resonance. The kinetic energy released in the dissociation depends on ramp speed and on the width of the resonance. This was previously used to determine the width of a fairly broad Feshbach resonance in Na \cite{mukaiyama:04}. We now used this method to determine the widths of four Feshbach resonances in $^{87}$Rb. We discuss, why the widths of narrow Feshbach resonances can be determined with this method even in the presence of significant magnetic-field noise. Indeed, three of the resonances investigated here are so narrow that their widths are not accessible otherwise. In addition, an outgoing quasi-mono-energetic atomic wave is created by jumping the magnetic field across the resonance instead of ramping linearly in time. In this case, the atoms fly apart on the surface of a hollow sphere.

We start with a brief summary of the theoretical background \cite{mukaiyama:04,goral:cond-mat0312178,yurovsky:cond-mat/0402334}. A Feshbach resonance arises when a closed-channel bound state crosses an open-channel dissociation threshold when varying the magnetic field. Near this crossing, the bare states are coupled to dressed states. The energy $\epsilon$ of the bare molecular state with respect to the dissociation threshold depends to a good approximation linearly on the magnetic field $B$ so that $\epsilon = (B - B_{\rm res}) \Delta\mu$, where the slope $\Delta\mu$ is the difference in the magnetic moments between a molecule and a dissociated atom pair, and $B_{\rm res}$ is the position of the Feshbach resonance.

A long-lived molecule can only exist for $\epsilon < 0$. Otherwise, the molecule dissociates spontaneously into a pair of unbound atoms due to the resonant coupling to the continuum of atomic pair states above the dissociation threshold. When a molecule at rest with $\epsilon > 0$ dissociates, the energy $\epsilon$ is released as kinetic energy of the two atoms. Since total momentum is conserved and both atoms have equal mass, the atoms have equal kinetic energy ($E=\epsilon/2$ each) and precisely opposite momentum vectors. The decay rate $\Gamma_{\rm mol}$ of the molecules can be obtained from Fermi's golden rule, yielding \cite{mukaiyama:04}
\begin{eqnarray}
 \label{eq-Gamma-mol}
\Gamma_{\rm mol}(\epsilon) = \frac{2 |\Delta B \Delta \mu|}{\hbar} \left( \frac{m\; a_{\rm bg}^2}{\hbar^2} \; \epsilon \right)^{1/2}
\end{eqnarray}
for $\epsilon >0$, and $\Gamma_{\rm mol}=0$ otherwise. Here, $\Delta B$ is the width of the Feshbach resonance, $m$ is the mass of an atom, and $a_{\rm bg}$ is the background value of the $s$-wave scattering length. Alternatively, Eq.~(\ref{eq-Gamma-mol}) can be obtained from scattering theory, see e.g.\ Eqs.~(25) and (42) in Ref.~\cite{timmermans:99} with $\Gamma_m=\hbar \Gamma_{\rm mol}$.

Consider a magnetic-field ramp that is linear in time with $d\epsilon/dt > 0$. In this case, the molecule fraction $f$ decays as \cite{mukaiyama:04}
\begin{eqnarray}
 \label{eq-molecule-fraction}
f(\epsilon) = \exp\left( - \; \frac{4 |\Delta B|}{3 \hbar \left| \frac{dB}{dt} \right| }
\; \sqrt{\frac{m a_{\rm bg}^2}{\hbar^2}} \; \epsilon^{3/2} \right) 
\end{eqnarray}
for $\epsilon >0$, and $f=1$ otherwise. The mean kinetic energy of a single atom after the dissociation is then \cite{mukaiyama:04,goral:cond-mat0312178}
\begin{eqnarray}
 \label{eq-mean-energy}
\langle E \rangle =  \frac{1}{2} \; \Gamma\left(\frac{5}{3}\right) 
  \left( \frac{3\hbar \left| \frac{dB}{dt} \right|}{4 |\Delta B|} \; \sqrt{\frac{\hbar^2}{m a_{\rm bg}^2}} \; \right)^{2/3} \; ,
\end{eqnarray}
where $\Gamma$ is the Euler gamma function with $\Gamma\left(\frac{5}{3}\right) \approx 0.903$. This relation makes it possible to determine the width $\Delta B$ of a Feshbach resonance from a measurement of $\langle E \rangle$, because $a_{\rm bg}$ is typically known with much better accuracy than $\Delta B$. Unlike previous methods, this method of measuring $\Delta B$ does not depend on knowledge of the initial atomic density distribution.

The probability density $D$ for the velocity $\vec v$ of the dissociated atoms is easily obtained from Eq.~(\ref{eq-molecule-fraction}), yielding
\begin{eqnarray}
 \label{eq-velocity-distribution}
 D(\vec v \,) d^3v 
 = \frac{3}{4 \pi v_0^3} \exp\left( - \frac{|\vec v\, |^3}{v_0^3}  \right)  \; d^3v \; ,
\end{eqnarray}
where $\langle E \rangle = \frac{m}{2} v_0^2 \Gamma\left(\frac{5}{3}\right)$. Using a time-of-flight method, this velocity distribution is converted into a position distribution, which is measured using absorption imaging.

In the experiment, the velocity distribution of the molecules before dissociation has a finite width. The result is an offset energy $E_0$ in Eq.~(\ref{eq-mean-energy}), but the shape of the velocity distribution Eq.~(\ref{eq-velocity-distribution}) is hardly affected, because $E_0$ is small.

A fit of Eq.~(\ref{eq-velocity-distribution}) to the absorption images would be cumbersome, because no analytic solution is known for the integral of Eq.~(\ref{eq-velocity-distribution}) along one or two coordinate axes. We therefore fit a two-dimensional Gaussian to the absorption images and extract $\langle E \rangle$ of the Gaussian fit function. In order to analyze the error made by this choice of the fit function, we numerically integrate Eq.~(\ref{eq-velocity-distribution}) along one coordinate axis for a specific value of $\langle E \rangle$ and fit a two-dimensional Gaussian to this. The resulting best-fit value $\langle E \rangle$ of the Gaussian is a factor of $\sim 1.18$ larger than the true value $\langle E \rangle$ of Eq.~(\ref{eq-velocity-distribution}). We correct the kinetic energies in our data analysis by this factor.

An experimental cycle begins with the production of a BEC of $^{87}$Rb atoms in a magnetic trap. The BEC is then transferred to an optical dipole trap and a magnetic field of up to $\sim 1000$~G is applied. The atomic spin is prepared in the absolute ground state $|f,m_f\rangle=|1,1\rangle$. This state has many Feshbach resonances \cite{marte:02}.

The dipole trap consists of two beams from a Nd:YAG laser that cross at right angels. One beam is horizontal, the other subtends an angle of $25^\circ$ with the horizontal plane. The horizontal beam has a beam waist ($1/e^2$ radius of intensity) of $40~\mu$m and a power of 45~mW, the other beam has $55~\mu$m and 20~mW. Trap frequencies of $2 \pi \times (80, 110, 170) $~Hz were measured with parametric heating.

In our previous experiments \cite{volz:03,duerr:04}, an undesired reflection of the horizontal beam from an uncoated inside surface of the glass cell housing the BEC caused a weak standing wave. This created a one-dimensional optical lattice with a well-depths of $\sim k_B\times 0.6~\mu$K, which exceeds typical values of $k_B T$ and of the chemical potential. The horizontal beam is now tilted by a few degrees with respect to the glass surface, thus eliminating this standing wave. This improves the trap loading substantially, because atoms from the wings of the trap are now free to move to the central region. Now, a BEC of typically $6\times 10^5$ atoms can be held in the crossed dipole trap, with a small thermal fraction of less than $1\times 10^5$ atoms. The in-trap peak density in the BEC is typically $5\times 10^{14}$~cm$^{-3}$.

\begin{figure} [bt]
\includegraphics[width=.4\textwidth]{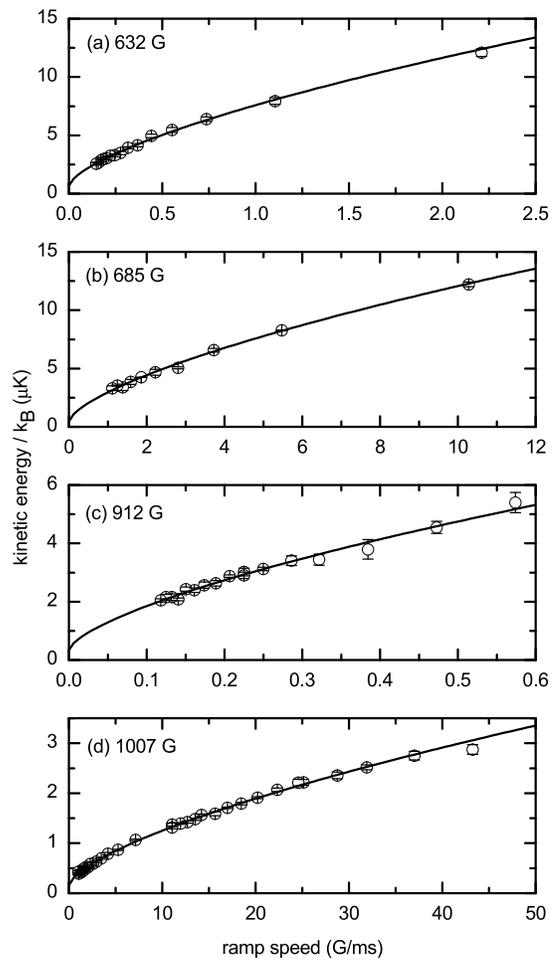}
\caption{\label{fig-energy}
Mean kinetic energy per atom as a function of the speed of the dissociation ramp. Parts (a), (b), (c), and (d) were measured at Feshbach resonances at 632~G, 685~G, 912~G, and 1007~G, respectively. The solid lines show fits of Eq.~(\ref{eq-mean-energy}) to the data. The best-fit values are shown in Tab.~\ref{tab-widths}.
 }
\end{figure}

Ultracold molecules are created by ramping the magnetic field slowly downward through one of the Feshbach resonances, as described in Ref.~\cite{duerr:04}. About 2~ms before the molecule creation, the atoms are released from the dipole trap. The fraction of the population that is converted into molecules is $\sim 10$\% for the broad resonances at 685~G and 1007~G, and it is $\sim 3$\% for the narrower resonances at 632~G and 912~G. We could not detect molecules when working at even narrower resonances with a width predicted to be $\Delta B \sim 0.2$~mG. We speculate that this is because the slow molecule-creation ramp might suffer from magnetic-field noise. After the molecule creation, a Stern-Gerlach field is applied to separate the molecules from the remaining atoms. Next, the magnetic field is ramped upward through the Feshbach resonance to dissociate the molecules. Then the atoms fly freely for up to 11~ms, before an absorption image of the cloud is taken on a video camera. The mean kinetic energy is extracted from the image as described above.

Experimental results as a function of ramp speed are shown in Fig.~\ref{fig-energy} for the 4 broadest Feshbach resonances of the state $|1,1\rangle$. Solid lines show fits of Eq.~(\ref{eq-mean-energy}) to the data, with the width of the resonance $\Delta B$ and the offset energy $E_0$ as free fit parameters. The best-fit values are shown in Tab.~\ref{tab-widths}, using the theory value $a_{\rm bg} =100.5$ Bohr radii for the state $|1,1\rangle$ from Ref.~\cite{volz:03}. For the resonances at 632~G, 912~G, and 1007~G, the measured widths agree well with the theoretical predictions and with a previously measured value, both also shown in Tab.~\ref{tab-widths}. The good agreement shows that the dissociation of molecules can be used as a reliable method to determine $\Delta B$. The resonance at 685~G, however, is a factor of $\sim 2.7$ narrower than predicted. More recent calculations \cite{vanKempen:pers} predict a width of 10~mG for this resonance, which is much closer to the value measured here.

\begin{table} [bt] 
\caption{
\label{tab-widths}
Position $B_{\rm res}$ and width $\Delta B$ of the Feshbach resonances. $\Delta B_{\rm fit}$ is the best-fit value obtained from the measurement in Fig.~\ref{fig-energy}. $\Delta B_{\rm th}$ is the theoretical prediction from Ref.~\cite{marte:02}. $\Delta B_{\rm prev}$ is the result of a previous measurement in Ref.~\cite{volz:03}. 
}
\begin{tabular}{cccc}
\hline
\hline
$B_{\rm res}$ (G) & $\Delta B_{\rm fit}$ (mG) & $\Delta B_{\rm th}$ (mG) & $\Delta B_{\rm prev}$ (mG) \\
\hline 
632  &    1.3(2) &   1.5       & $-$     \\
685  &    6.2(6) &   17       & $-$     \\
912  &    1.3(2) &   1.3      & $-$     \\
1007 &    210(20) &   170(30)  & 200(30) \\
\hline
\hline
\end{tabular}
\end{table}

It is surprising that a width as small as $\Delta B = 1.3$~mG can be measured with our setup, because the magnetic-field noise is most likely larger than this value. The observed linewidths of microwave transitions measured with 50~ms long pulses sets an experimental upper bound on magnetic-field noise of 4~mG (rms). An attempt to directly measure the magnetic-field dependence of the scattering length $a(B)$ for the 632~G or 912~G resonance would therefore most likely suffer strongly from the magnetic-field noise.

There are two reasons, why such a small $\Delta B$ can be measured with the method presented here nonetheless. First, the dissociation process is pretty fast, e.g., $\Gamma_{\rm mol} \sim 10$~kHz for $\epsilon = k_B \times 5~\mu$K at the 912~G resonance. Therefore, low-frequency magnetic-field noise merely shifts the exact time of dissociation but has no effect on the actual ramp speed during the rather short decay time. Second, by choosing fast enough ramp speeds, the experiment is operated in a regime where the relevant decay happens at magnetic fields that are pretty far away from $B_{\rm res}$. A typical decay energy of $\epsilon = k_B \times 5~\mu$K corresponds to $B-B_{\rm res} \sim 25$~mG for the 912~G resonance. Here, the magnetic-field noise has little effect on the molecule-decay rate.

In addition to the possibility to measure the widths of Feshbach resonances, the dissociation of molecules into atom pairs can also be used to produce a mono-energetic spherical wave of atoms. To this end, the magnetic-field is jumped across the resonance as fast as possible and then held at a fixed value $B_{\rm final}$. If the jump is fast enough, there will be hardly any dissociation during the jump. Therefore all molecules dissociate at $B_{\rm final}$, thus creating atoms with a fixed amount of kinetic energy. Hence, the atoms fly apart on the surface of a hollow sphere during the subsequent free flight.

\begin{figure} [bt]
\includegraphics[width=.3\textwidth]{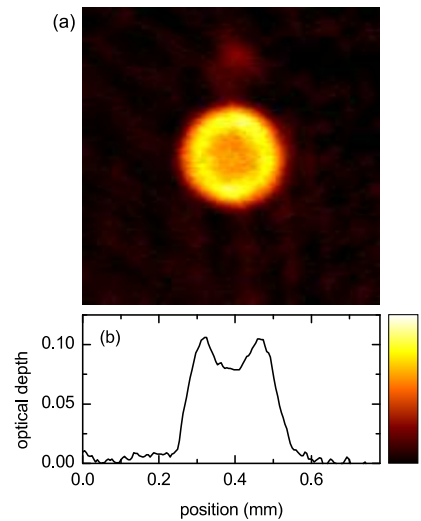}
\caption{\label{fig-ring}
(color online) (a) Mono-energetic spherical wave of atoms, created by dissociation of molecules when jumping the magnetic field across the Feshbach resonance. The atoms fly apart on the surface of a hollow sphere. The absorption-imaging beam integrates the three-dimensional density profile along one direction. The observed two-dimensional image therefore shows a ring, with a non-zero density in the center. The image was averaged over $\sim 100$ experimental shots. (b) Line profile across the center of the image. The dip in the center is clearly visible.
 }
\end{figure}

The laser beam used in absorption imaging integrates this density distribution along its propagation direction, thus yielding a two-dimensional image that shows a ring, with a non-zero density in the center. Such an image is shown for the 685~G resonance in Fig.~\ref{fig-ring}a. Due to the initial momentum spread of the molecules the dissociated atoms are not perfectly mono-energetic. This smears the atomic distribution around the ring. The contrast is still good enough to see a clear dip in the center of the line profile in Fig.~\ref{fig-ring}b. The data were taken with 5.5~ms time-of-flight after the dissociation and with $B_{\rm final} - B_{\rm res} \sim 40$~mG corresponding to $\Gamma_{\rm mol} \sim 20$~kHz.

Technically, the creation of a sharp corner between the magnetic-field jump and the plateau at $B_{\rm final}$ is difficult. The experimental requirements concerning the sharpness of this corner are more stringent with broader resonances because of the faster molecule-decay rates. We believe that this is why we were unable to observe a dip as in Fig.~\ref{fig-ring}b at the 1007~G resonance with the given bandwidth of the servo that controls our magnetic field.

The scheme with the jump in the magnetic field can be easily generalized by using an arbitrary shape of the dissociation ramp, which makes it possible to tailor the time and energy dependence of the outgoing atomic wave function.

As mentioned earlier, one expects that after the dissociation of molecules the created atom pairs should have precisely opposite momentum vectors, given that the molecules had a negligible momentum spread initially. A detection of this pair correlation might be an interesting subject of future investigations.

In conclusion, dissociation of molecules with a linear magnetic-field ramp is a simple and reliable technique to determine the widths of Feshbach resonances, even in the presence of magnetic-field noise. We measured the widths of 4 Feshbach resonances in $^{87}$Rb thereby covering more than 2 orders of magnitude in $\Delta B$. We also showed that dissociation of molecules can be used to create a quasi-mono-energetic atomic wave.

We thank Cheng Chin for pointing out the possibility of creating mono-energetic atoms by jumping the magnetic field across the resonance. We also acknowledge fruitful discussions with Thorsten K{\"o}hler and Paul Julienne. This work was supported by the European-Union Network ``Cold Quantum Gases".


\begin{thebibliography}{20}
\expandafter\ifx\csname natexlab\endcsname\relax\def\natexlab#1{#1}\fi
\expandafter\ifx\csname bibnamefont\endcsname\relax
  \def\bibnamefont#1{#1}\fi
\expandafter\ifx\csname bibfnamefont\endcsname\relax
  \def\bibfnamefont#1{#1}\fi
\expandafter\ifx\csname citenamefont\endcsname\relax
  \def\citenamefont#1{#1}\fi
\expandafter\ifx\csname url\endcsname\relax
  \def\url#1{\texttt{#1}}\fi
\expandafter\ifx\csname urlprefix\endcsname\relax\def\urlprefix{URL }\fi
\providecommand{\bibinfo}[2]{#2}
\providecommand{\eprint}[2][]{\url{#2}}

\bibitem[{\citenamefont{Greiner et~al.}(2003)\citenamefont{Greiner, Regal, and
  Jin}}]{greiner:03}
\bibinfo{author}{\bibfnamefont{M.}~\bibnamefont{Greiner}},
  \bibinfo{author}{\bibfnamefont{C.~A.} \bibnamefont{Regal}}, \bibnamefont{and}
  \bibinfo{author}{\bibfnamefont{D.~S.} \bibnamefont{Jin}},
  \bibinfo{journal}{Nature (London)} \textbf{\bibinfo{volume}{426}},
  \bibinfo{pages}{537} (\bibinfo{year}{2003}).

\bibitem[{\citenamefont{Jochim et~al.}(2003)\citenamefont{Jochim, Bartenstein,
  Altmeyer, Hendl, Riedl, Chin, Hecker-Denschlag, and Grimm}}]{jochim:03}
\bibinfo{author}{\bibfnamefont{S.}~\bibnamefont{Jochim}},
  \bibinfo{author}{\bibfnamefont{M.}~\bibnamefont{Bartenstein}},
  \bibinfo{author}{\bibfnamefont{A.}~\bibnamefont{Altmeyer}},
  \bibinfo{author}{\bibfnamefont{G.}~\bibnamefont{Hendl}},
  \bibinfo{author}{\bibfnamefont{S.}~\bibnamefont{Riedl}},
  \bibinfo{author}{\bibfnamefont{C.}~\bibnamefont{Chin}},
  \bibinfo{author}{\bibfnamefont{J.}~\bibnamefont{Hecker-Denschlag}},
  \bibnamefont{and} \bibinfo{author}{\bibfnamefont{R.}~\bibnamefont{Grimm}},
  \bibinfo{journal}{Science} \textbf{\bibinfo{volume}{302}},
  \bibinfo{pages}{2101} (\bibinfo{year}{2003}).

\bibitem[{\citenamefont{Zwierlein et~al.}(2003)\citenamefont{Zwierlein, Stan,
  Schunck, Raupach, Gupta, Hadzibabic, and Ketterle}}]{zwierlein:03}
\bibinfo{author}{\bibfnamefont{M.~W.} \bibnamefont{Zwierlein}},
  \bibinfo{author}{\bibfnamefont{C.~A.} \bibnamefont{Stan}},
  \bibinfo{author}{\bibfnamefont{C.~H.} \bibnamefont{Schunck}},
  \bibinfo{author}{\bibfnamefont{S.~M.} \bibnamefont{Raupach}},
  \bibinfo{author}{\bibfnamefont{S.}~\bibnamefont{Gupta}},
  \bibinfo{author}{\bibfnamefont{Z.}~\bibnamefont{Hadzibabic}},
  \bibnamefont{and} \bibinfo{author}{\bibfnamefont{W.}~\bibnamefont{Ketterle}},
  \bibinfo{journal}{Phys. Rev. Lett.} \textbf{\bibinfo{volume}{91}},
  \bibinfo{pages}{250401} (\bibinfo{year}{2003}).

\bibitem[{bou()}]{bourdel:cond-mat/0403091}
\emph{\bibinfo{title}{{\rm T. Bourdel, L. Khaykovich, J Cubizolles, J. Zhang,
  F. Chevy, M. Teichmann, L. Tarruell, S.~J. Kokkelmans, and C. Salomon,
  e-print cond-mat/0403091}}}.

\bibitem[{\citenamefont{Regal et~al.}(2004)\citenamefont{Regal, Greiner, and
  Jin}}]{regal:04}
\bibinfo{author}{\bibfnamefont{C.~A.} \bibnamefont{Regal}},
  \bibinfo{author}{\bibfnamefont{M.}~\bibnamefont{Greiner}}, \bibnamefont{and}
  \bibinfo{author}{\bibfnamefont{D.~S.} \bibnamefont{Jin}},
  \bibinfo{journal}{Phys. Rev. Lett.} \textbf{\bibinfo{volume}{92}},
  \bibinfo{pages}{040403} (\bibinfo{year}{2004}).

\bibitem[{\citenamefont{Bartenstein et~al.}(2004)\citenamefont{Bartenstein,
  Altmeyer, Riedl, Jochim, Chin, Hecker-Denschlag, and Grimm}}]{bartenstein:04}
\bibinfo{author}{\bibfnamefont{M.}~\bibnamefont{Bartenstein}},
  \bibinfo{author}{\bibfnamefont{A.}~\bibnamefont{Altmeyer}},
  \bibinfo{author}{\bibfnamefont{S.}~\bibnamefont{Riedl}},
  \bibinfo{author}{\bibfnamefont{S.}~\bibnamefont{Jochim}},
  \bibinfo{author}{\bibfnamefont{C.}~\bibnamefont{Chin}},
  \bibinfo{author}{\bibfnamefont{J.}~\bibnamefont{Hecker-Denschlag}},
  \bibnamefont{and} \bibinfo{author}{\bibfnamefont{R.}~\bibnamefont{Grimm}},
  \bibinfo{journal}{Phys. Rev. Lett.} \textbf{\bibinfo{volume}{92}},
  \bibinfo{pages}{120401} (\bibinfo{year}{2004}).

\bibitem[{\citenamefont{Zwierlein et~al.}(2004)\citenamefont{Zwierlein, Stan,
  Schunck, Raupach, Kerman, and Ketterle}}]{zwierlein:04}
\bibinfo{author}{\bibfnamefont{M.~W.} \bibnamefont{Zwierlein}},
  \bibinfo{author}{\bibfnamefont{C.~A.} \bibnamefont{Stan}},
  \bibinfo{author}{\bibfnamefont{C.~H.} \bibnamefont{Schunck}},
  \bibinfo{author}{\bibfnamefont{S.~M.} \bibnamefont{Raupach}},
  \bibinfo{author}{\bibfnamefont{A.~J.} \bibnamefont{Kerman}},
  \bibnamefont{and} \bibinfo{author}{\bibfnamefont{W.}~\bibnamefont{Ketterle}},
  \bibinfo{journal}{Phys. Rev. Lett.} \textbf{\bibinfo{volume}{92}},
  \bibinfo{pages}{120403} (\bibinfo{year}{2004}).

\bibitem[{\citenamefont{Kinast et~al.}(2004)\citenamefont{Kinast, Hemmer, Gehm,
  Turlapov, and Thomas}}]{kinast:04}
\bibinfo{author}{\bibfnamefont{J.}~\bibnamefont{Kinast}},
  \bibinfo{author}{\bibfnamefont{S.~L.} \bibnamefont{Hemmer}},
  \bibinfo{author}{\bibfnamefont{M.~E.} \bibnamefont{Gehm}},
  \bibinfo{author}{\bibfnamefont{A.}~\bibnamefont{Turlapov}}, \bibnamefont{and}
  \bibinfo{author}{\bibfnamefont{J.~E.} \bibnamefont{Thomas}},
  \bibinfo{journal}{Phys. Rev. Lett.} \textbf{\bibinfo{volume}{92}},
  \bibinfo{pages}{150402} (\bibinfo{year}{2004}).

\bibitem[{\citenamefont{Strecker et~al.}(2003)\citenamefont{Strecker,
  Partridge, and Hulet}}]{strecker:03}
\bibinfo{author}{\bibfnamefont{K.~E.} \bibnamefont{Strecker}},
  \bibinfo{author}{\bibfnamefont{G.~B.} \bibnamefont{Partridge}},
  \bibnamefont{and} \bibinfo{author}{\bibfnamefont{R.~G.} \bibnamefont{Hulet}},
  \bibinfo{journal}{Phys. Rev. Lett.} \textbf{\bibinfo{volume}{91}},
  \bibinfo{pages}{080406} (\bibinfo{year}{2003}).

\bibitem[{\citenamefont{Donley et~al.}(2002)\citenamefont{Donley, Claussen,
  Thomson, and Wieman}}]{donley:02}
\bibinfo{author}{\bibfnamefont{E.~A.} \bibnamefont{Donley}},
  \bibinfo{author}{\bibfnamefont{N.~R.} \bibnamefont{Claussen}},
  \bibinfo{author}{\bibfnamefont{S.~T.} \bibnamefont{Thomson}},
  \bibnamefont{and} \bibinfo{author}{\bibfnamefont{C.~E.}
  \bibnamefont{Wieman}}, \bibinfo{journal}{Nature (London)}
  \textbf{\bibinfo{volume}{417}}, \bibinfo{pages}{529} (\bibinfo{year}{2002}).

\bibitem[{\citenamefont{Herbig et~al.}(2003)\citenamefont{Herbig, Kraemer,
  Mark, Weber, Chin, N{\"a}gerl, and Grimm}}]{herbig:03}
\bibinfo{author}{\bibfnamefont{J.}~\bibnamefont{Herbig}},
  \bibinfo{author}{\bibfnamefont{T.}~\bibnamefont{Kraemer}},
  \bibinfo{author}{\bibfnamefont{M.}~\bibnamefont{Mark}},
  \bibinfo{author}{\bibfnamefont{T.}~\bibnamefont{Weber}},
  \bibinfo{author}{\bibfnamefont{C.}~\bibnamefont{Chin}},
  \bibinfo{author}{\bibfnamefont{H.-C.} \bibnamefont{N{\"a}gerl}},
  \bibnamefont{and} \bibinfo{author}{\bibfnamefont{R.}~\bibnamefont{Grimm}},
  \bibinfo{journal}{Science} \textbf{\bibinfo{volume}{301}},
  \bibinfo{pages}{1510} (\bibinfo{year}{2003}).

\bibitem[{\citenamefont{D{\"u}rr et~al.}(2004)\citenamefont{D{\"u}rr, Volz,
  Marte, and Rempe}}]{duerr:04}
\bibinfo{author}{\bibfnamefont{S.}~\bibnamefont{D{\"u}rr}},
  \bibinfo{author}{\bibfnamefont{T.}~\bibnamefont{Volz}},
  \bibinfo{author}{\bibfnamefont{A.}~\bibnamefont{Marte}}, \bibnamefont{and}
  \bibinfo{author}{\bibfnamefont{G.}~\bibnamefont{Rempe}},
  \bibinfo{journal}{Phys. Rev. Lett.} \textbf{\bibinfo{volume}{92}},
  \bibinfo{pages}{020406} (\bibinfo{year}{2004}).

\bibitem[{\citenamefont{Xu et~al.}(2003)\citenamefont{Xu, Mukaiyama,
  Abo-Shaeer, Chin, Miller, and Ketterle}}]{xu:03}
\bibinfo{author}{\bibfnamefont{K.}~\bibnamefont{Xu}},
  \bibinfo{author}{\bibfnamefont{T.}~\bibnamefont{Mukaiyama}},
  \bibinfo{author}{\bibfnamefont{J.~R.} \bibnamefont{Abo-Shaeer}},
  \bibinfo{author}{\bibfnamefont{J.~K.} \bibnamefont{Chin}},
  \bibinfo{author}{\bibfnamefont{D.~E.} \bibnamefont{Miller}},
  \bibnamefont{and} \bibinfo{author}{\bibfnamefont{W.}~\bibnamefont{Ketterle}},
  \bibinfo{journal}{Phys. Rev. Lett.} \textbf{\bibinfo{volume}{91}},
  \bibinfo{pages}{210402} (\bibinfo{year}{2003}).

\bibitem[{\citenamefont{Mukaiyama et~al.}(2004)\citenamefont{Mukaiyama,
  Abo-Shaeer, Xu, Chin, and Ketterle}}]{mukaiyama:04}
\bibinfo{author}{\bibfnamefont{T.}~\bibnamefont{Mukaiyama}},
  \bibinfo{author}{\bibfnamefont{J.~R.} \bibnamefont{Abo-Shaeer}},
  \bibinfo{author}{\bibfnamefont{K.}~\bibnamefont{Xu}},
  \bibinfo{author}{\bibfnamefont{J.~K.} \bibnamefont{Chin}}, \bibnamefont{and}
  \bibinfo{author}{\bibfnamefont{W.}~\bibnamefont{Ketterle}},
  \bibinfo{journal}{Phys. Rev. Lett.} \textbf{\bibinfo{volume}{92}},
  \bibinfo{pages}{180402} (\bibinfo{year}{2004}).

\bibitem[{gor()}]{goral:cond-mat0312178}
\emph{\bibinfo{title}{{\rm K. Goral, T. K{\"o}hler, S. A. Gardiner, E.
  Tiesinga, and P. Julienne, e-print cond-mat/0312178}}}.

\bibitem[{yur()}]{yurovsky:cond-mat/0402334}
\emph{\bibinfo{title}{{\rm V. A. Yurovsky and A. Ben-Reuven, e-print
  cond-mat/0402334}}}.

\bibitem[{\citenamefont{Timmermans et~al.}(1999)\citenamefont{Timmermans,
  Tommasini, Hussein, and Kerman}}]{timmermans:99}
\bibinfo{author}{\bibfnamefont{E.}~\bibnamefont{Timmermans}},
  \bibinfo{author}{\bibfnamefont{P.}~\bibnamefont{Tommasini}},
  \bibinfo{author}{\bibfnamefont{M.}~\bibnamefont{Hussein}}, \bibnamefont{and}
  \bibinfo{author}{\bibfnamefont{A.}~\bibnamefont{Kerman}},
  \bibinfo{journal}{Phys. Rep.} \textbf{\bibinfo{volume}{315}},
  \bibinfo{pages}{199} (\bibinfo{year}{1999}).

\bibitem[{\citenamefont{Marte et~al.}(2002)\citenamefont{Marte, Volz, Schuster,
  D{\"u}rr, Rempe, van Kempen, and Verhaar}}]{marte:02}
\bibinfo{author}{\bibfnamefont{A.}~\bibnamefont{Marte}},
  \bibinfo{author}{\bibfnamefont{T.}~\bibnamefont{Volz}},
  \bibinfo{author}{\bibfnamefont{J.}~\bibnamefont{Schuster}},
  \bibinfo{author}{\bibfnamefont{S.}~\bibnamefont{D{\"u}rr}},
  \bibinfo{author}{\bibfnamefont{G.}~\bibnamefont{Rempe}},
  \bibinfo{author}{\bibfnamefont{E.~G.} \bibnamefont{van Kempen}},
  \bibnamefont{and} \bibinfo{author}{\bibfnamefont{B.~J.}
  \bibnamefont{Verhaar}}, \bibinfo{journal}{Phys. Rev. Lett.}
  \textbf{\bibinfo{volume}{89}}, \bibinfo{pages}{283202}
  (\bibinfo{year}{2002}).

\bibitem[{\citenamefont{Volz et~al.}(2003)\citenamefont{Volz, D{\"u}rr, Ernst,
  Marte, and Rempe}}]{volz:03}
\bibinfo{author}{\bibfnamefont{T.}~\bibnamefont{Volz}},
  \bibinfo{author}{\bibfnamefont{S.}~\bibnamefont{D{\"u}rr}},
  \bibinfo{author}{\bibfnamefont{S.}~\bibnamefont{Ernst}},
  \bibinfo{author}{\bibfnamefont{A.}~\bibnamefont{Marte}}, \bibnamefont{and}
  \bibinfo{author}{\bibfnamefont{G.}~\bibnamefont{Rempe}},
  \bibinfo{journal}{Phys. Rev. A} \textbf{\bibinfo{volume}{68}},
  \bibinfo{pages}{010702(R)} (\bibinfo{year}{2003}).

\bibitem[{van()}]{vanKempen:pers}
\emph{\bibinfo{title}{{\rm Eric van Kempen and Boudewijn Verhaar, personal
  communication.}}}

\end{thebibliography}

\end{document}